# Design Aesthetics Recommender System Based On Customer Profile And Wanted Affect

Brahim BENAISSA*[a], Masakazu KOBAYASHI [a], Keita KINOSHITA [a]

[a] Toyota Technological Institute, Department of Mechanical Systems Engineering, Design Engineering Lab, Japan, Corresponding author: benaissa@toyota-ti.ac.jp

**ABSTRACT**

Product recommendation systems have been instrumental in online commerce since the early days. Their development is expanded further with the help of big data and advanced deep learning methods, where consumer profiling is central. The interest of the consumer can now be predicted based on the personal past choices and the choices of similar consumers. However, what is currently defined as a choice is based on quantifiable data, like product features, cost, and type. This paper investigates the possibility of profiling customers based on the preferred product design and wanted affects. We considered the case of vase design, where we study individual Kansei of each design. The personal aspects of the consumer considered in this study were decided based on our literature review conclusions on the consumer response to product design. We build a representative consumer model that constitutes the recommendation system's core using deep learning. It asks the new consumers to provide what affect they are looking for, through Kansei adjectives, and recommend; as a result, the aesthetic design that will most likely cause that affect.

## 1 INTRODUCTION

Recommender systems have been crucial for online commerce, and their importance keeps growing with the increasing amount of products in the market. While the consumer profiling systems have effectively solved the problem of product recommendation, the problem of "design recommendation" is more challenging because an aesthetic design cannot be handled in a quantifiable manner, similar to the product features. Therefore, solving this problem requires recommender systems to handle design aesthetics and consider that consumer perception of design is variable.

Different consumer groups perceive the same design differently [1]. Research studies related to consumer perception of a product design suggested that we adopted a unified position in product aesthetics appreciation; thus, consumers from different backgrounds can perceive modern design similarly [2]. It was found that consumer basic profile like gender and age, is not significant in aesthetic perception; however, consumer personality, namely the aspect of openness to experiences, correlates with perception variation of aesthetic design [3]–[5].

Recommender systems were born out of necessity in the early days of the internet due to the exponentially growing amount of information. The earlier item recommendation systems were centered around consumer feedback, where early consumer rating is used to rank the content for new consumers [6]. Then, hybrid recommenders introduced the use of consumer action information, such as buying an item or consuming content, besides manual rating [7]. This idea encouraged collecting actions of multiple users, which led to more powerful recommenders based on consumer profiling [8] coupled with the use of item information [9]. Further development was made through context-aware recommender systems [10].

Several recommendation systems stem from these basic ideas, such as the group recommendation system, where a preference of a group member related to an item can help predict the position of all group members regarding that item [11]. Other recommendation systems assume that consumers with neighbouring characteristics most likely have similar preferences [12]. Or manually insert consumer preferences constraints in the consumer profile [13]. Personality is first introduced in such systems [14] to consider that consumer rating is highly dependent on the personality aspect of agreeableness [15]. Then later, to take advantage of the similarity of presences between closely positioned personalities in music [16], TV programs [17], video games [18], and products [19].

In Kansei studies, the consumer affect is considered the current emotion compound. It corresponds to the most abstract, positive or negative, responses to a product design [20], [21]. Machine learning approaches have been employed in multiple domains [22]–[27], including Kansei studies [28]–[36]. Where it has been used to model the affective response in the emerging field of aesthetic design recommendation [36], [37]. In this paper, we suggest a new method for aesthetic design recommendation based on consumer openness traits and wanted affect. The article is organized as follows; first, we present the proposed aesthetic recommendation system, then the implementation procedure for consumer affect modelling, describe the consumer affect experiment, and discuss the results.

## 2   PROPOSED AESTHETIC DESIGN RECOMMENDER SYSTEM

Figure 1. Illustrate the proposed aesthetic design recommendation system (ADRS) based on multiple affective responses and wanted affect. It consists of two parts: the first part is the offline modelling of the affective response to a set of variable designs collected from various consumers. And the second part consists of the interactive recommender, which will collect the personal aspects of an individual consumer and the affect they are looking for from a design and suggest the design that will most likely fit the affect wanted by this particular individual.

In a consumer modelling problem, an individual can be represented by multiple variables such as age, gender, work, etc. However, in our research review [5], we found that, when it comes to product design evaluation, only a few parameters have an observable connection with the aesthetic evaluation of a design. Namely, the personality aspect of openness, prior exposure to similar designs, and mood at the moment of evaluation. So in this study, we attempt to model an individual consumer based on this information.

On the other hand, we collect the affective response to a set of designs using affective adjectives. Figure 2. Illustrate the research framework of the proposed method where we select the design variables of interest and create product variations. We then collect the affective responses of various consumers, using a set of designs that cover the scope of design variables. At the same time, we collect the consumer's characteristics. Using a Deep Neural network, we build the model mapping the connection between the individual character of the consumers and their affective responses. And used to predict the affective reaction of new consumers, using personal characteristics as input.

The recommender system then uses the consumer's wanted affect as the reference and finds the product design variation corresponding to that affective response's maximum. This response can be different from one consumer to another because their predicted affective response can be different based on their characteristics.

## 3   IMPLEMENTATION PROCEDURES

We selected for this study the vase product for three reasons. First, it is a typical product that most people are familiar with and many own. Second, its aesthetics can be very distinguishable; therefore, variation in its design can evoke a wide range of affective responses. And lastly, decent designs can be generated mathematically, allowing flexibility in terms of automatic design generation and optimization based on consumer affective needs.

A vase here has two design variables: the size of the opening and the curvature. Plus two additional variables for texture, namely the number of vertical lines and the number of horizontal lines. The combination of these variables results in various aesthetic designs. Figure 3. Depicting the nine examples considered in the affective response experiment, they were chosen to represent diverse aesthetic looks, defining every design variable and at least treating each variable twice.

Participants were asked to rate the affective response of each design by selecting five degrees of affect (strongly disagree, disagree, neutral, agree, and strongly agree) in 12 affective adjectives, namely: Feminine, Emotional, Delicate, Elegant, Technological, Strong, Gentle, Traditional, Loud, Stable, Practical, and Luxurious. These Kansei adjectives were chosen based on research that studied vase design affective response and justified the relevance of these adjectives [1], [31], [38].

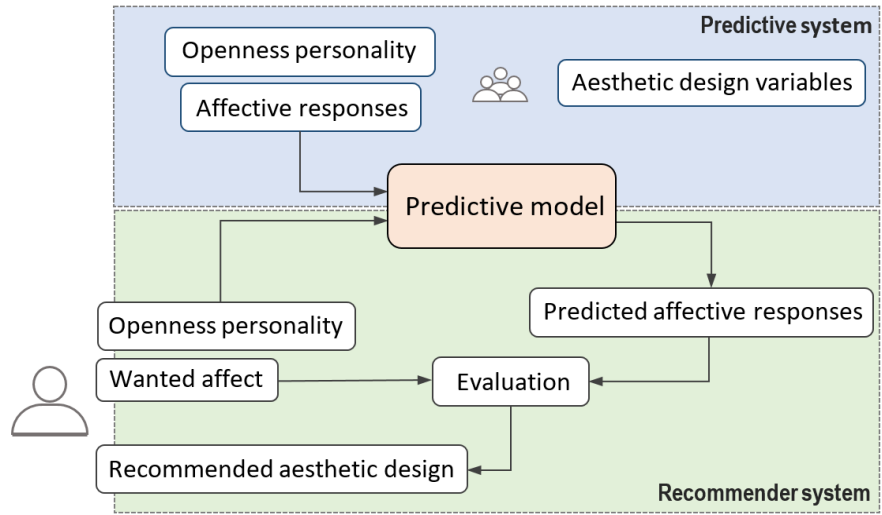

**Figure 1**. Overview of the affective response based ADRS

From the perspective of personal characteristics, we considered one question for mood, one question for exposure, and five questions for openness personality. These questions are as follows:

1. How is your day going?
   - Options: Terrific!, Good and Other.
2. Do you own a vase at home?
   - Options: Yes and No.
3. Is it fun to be in the museum?
4. Do you enjoy discussing new ideas?
5. Do you love adventure?
6. Are you excited to try new activities?
7. Do you avoid philosophical discussions?
   - Options in question 3 to question 7:  Not at all, Not much, A little and Very much.

We collect the affective response data using an online survey, where participants rated the pictures of the vase designs by selecting radio buttons corresponding to their affective responses. All participants rated the entirety of the design examples using all affective adjectives, resulting in 108 affective responses from each user. We collected 88 responses from participants of various backgrounds, ages, genders, and cultures, 41 participants were Japanese speakers, and 47 were English speakers.

Several research studies investigated the individual characteristics of the consumer but found a weak significance in consumer profile [1], intelligence [39],  and culture [40]. However, personality [41], [42] and exposure [43] were found to be the strongest individual factors influencing the consumer design response. To investigate that, we clustered the participants based on their personality characteristics using the K-means algorithm.

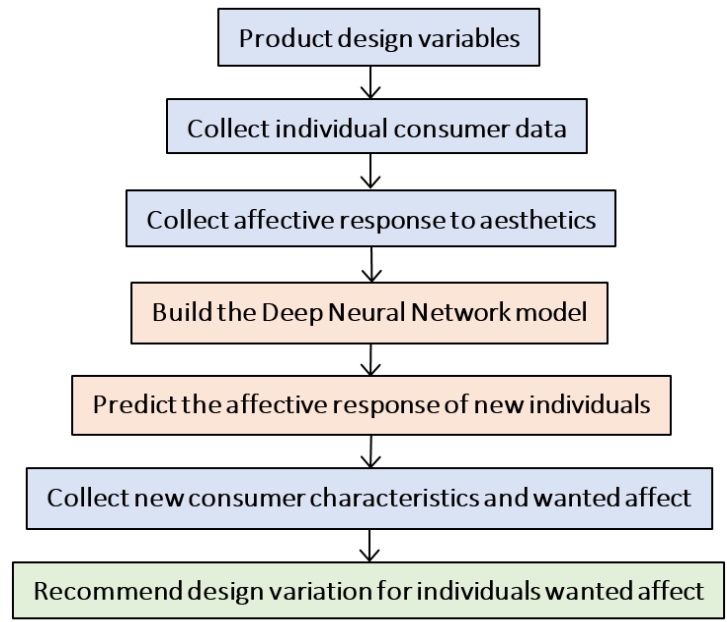

**Figure 2**. The research framework for the predictive model

The clustering result is considered after 100 algorithm runs. We found that the language did not significantly correlate with clusters, and each cluster contained equivalent members from the different language speakers. The influence of the questions on the first and second components is shown in figure 4. (a). The principle components analysis resulted in the contribution rates shown in 4. (b). For instance, we can observe the first component is influenced by most questions, with the first question being important for both principal components.

It is observed that questions 2 and 3, are majorly represented by the first component, while question 5 is represented by the second principal component. Considering the information collected from the questions, we accept that the first principal component represents the degree of openness to experience, while the second principal component indicates whether the tendency is physical or mental. The principal component scores for the participants are shown in Figures 5 and 6.

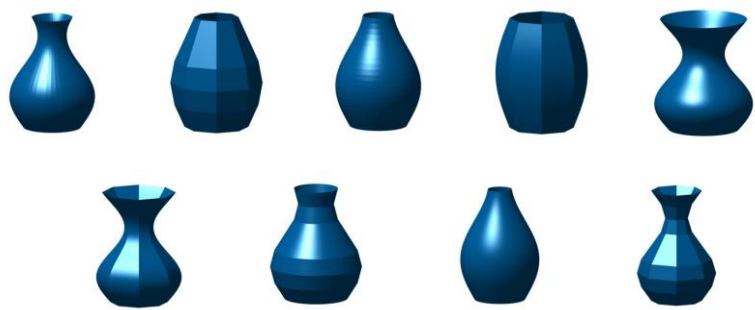

**Figure 3**. Representative Aesthetic design samples

Figure 5. shows the clustering result for the case of two clusters, where the first cluster contained a group of consumers with innovative and mental tendencies with 42 members. The second cluster grouped 33 participants with innovative and physical tendencies. Figure 6. shows the clustering result for the case of three clusters containing a group of consumers with conservative tendencies with 12 members. And split the participants of open tendencies into two subgroups. In the context of this study, it is determined that three clusters were appropriate.

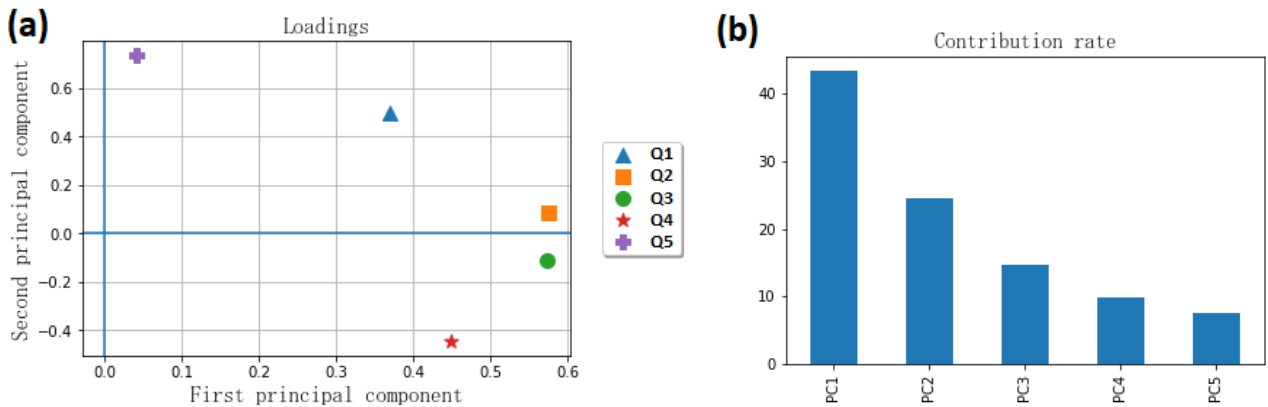

**Figure 4**. Principal Component Analysis

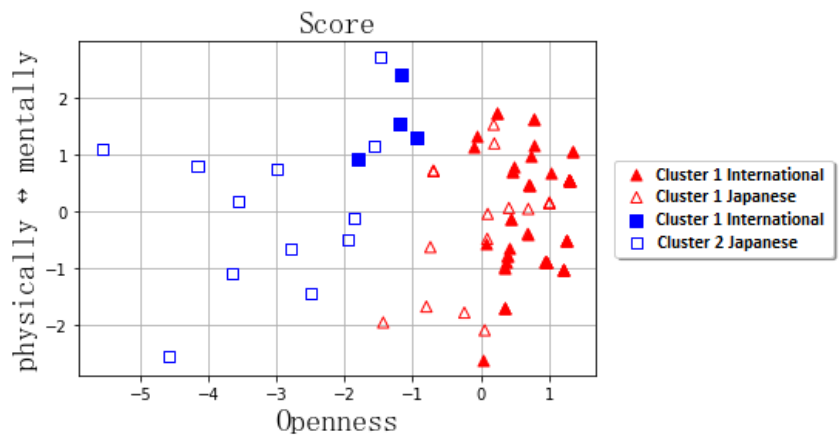

**Figure 5**. Two cultures study

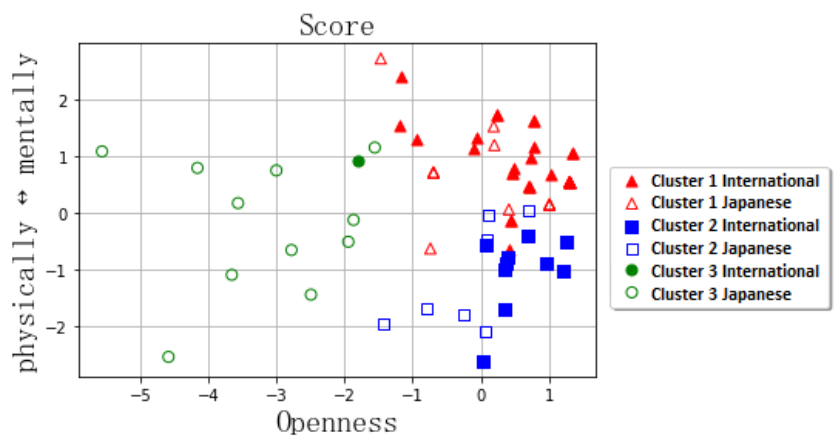

**Figure 6**. Three cultures study

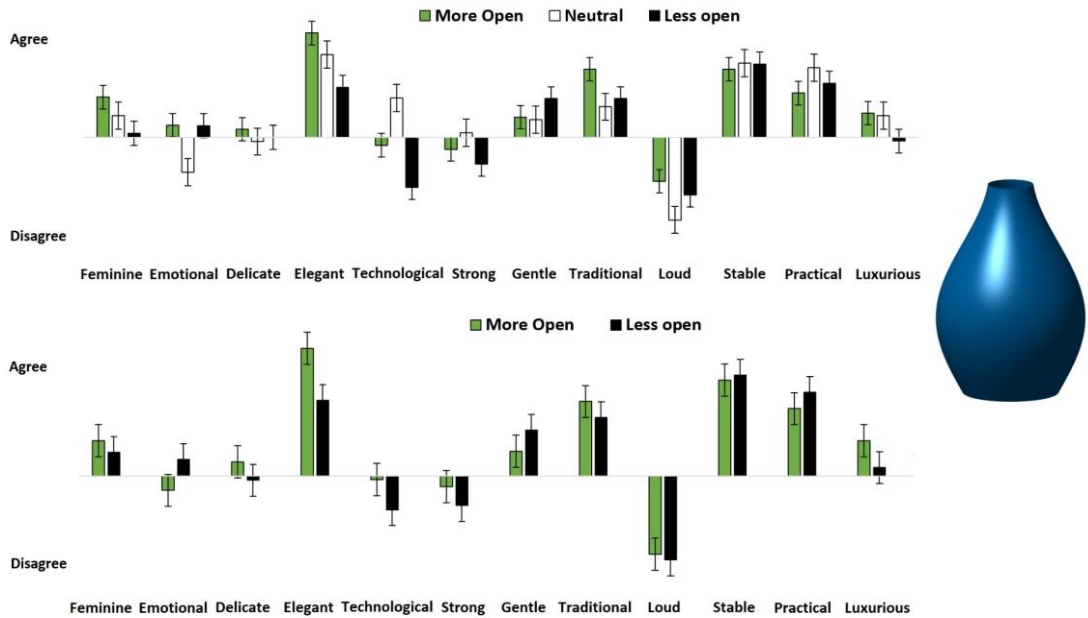

**Figure 7**. Average affective response for vase1, with 2 and 3 personal characteristic clusters

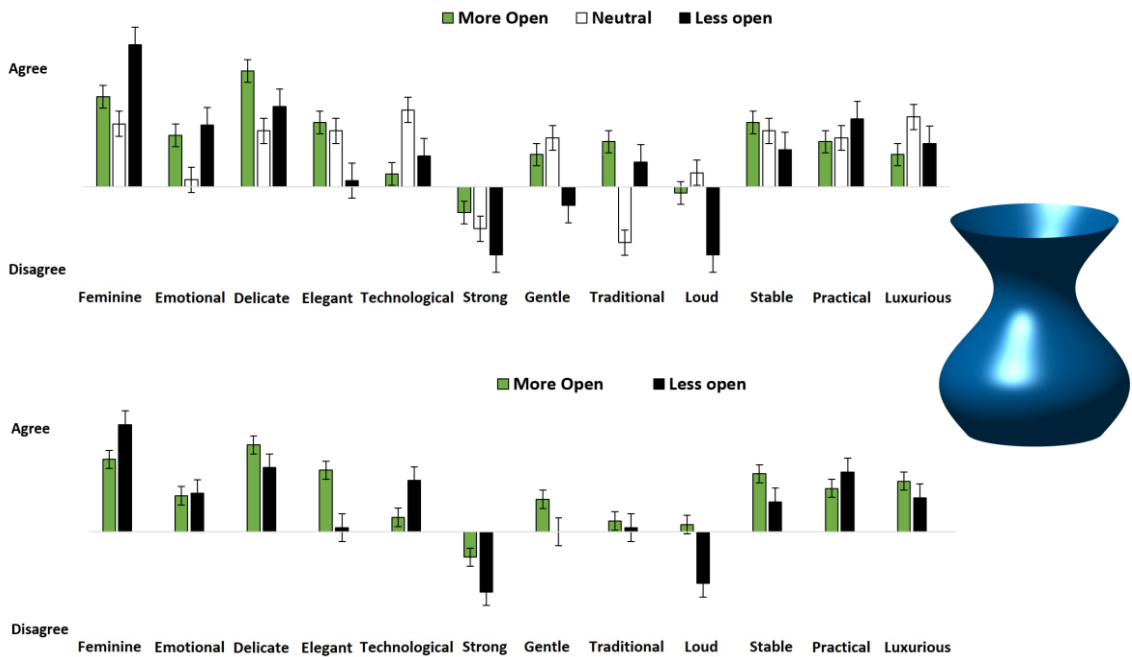

**Figure 8**. Average affective response for vase 5, with 2 and 3 personal characteristic clusters

Figure 7. Shows the average affective response for vase 1, with participants clustered in 2 and 3 clusters. And Figure 8. Shows the average affective response for vase 5, with participants clustered in 2 and 3 clusters. These figures show that the participants with more open personalities and lower open personalities have a similar affective response in most cases, but almost every time to a different degree. They have some disagreements, like in the case of "Emotional" in vase 1 and the case of "Loud" in vase 5. The cluster "Neutral personality" also has its differences from the later clusters, like in the case of "Emotional" in vase 1 and "Traditional" in vase 5. This response deference

is in line with the research findings and confirms the significance of the personality factor in the individual product design response.

## 4 RECOMMENDER SYSTEM RESULTS

To model the affective response of consumers, we use a feedforward Deep Neural network and conjugate gradient for network training. The network has eight hidden layers. And the recommendation algorithm is based on the idea of comparing the wanted affect with the predicted affect and outputting the design that corresponds to the highest value in that particular affect adjective.

In this experiment, we consider six testing participants and assume they are all looking for the same affects for the sake of consistency, namely "Feminine" and "Technological". The result of this experiment is presented in figure 9. It shows that this system can recommend different aesthetic designs for different people based on their personality characteristics. This system also allows recommending the same design for different desired affects of different individuals. Such as the recommended aesthetic designs for the 2$^{nd}$ and 5$^{th}$ individuals. And lastly, it may recommend the same design for different wanted aesthetic affects, like in the case of 1$^{st}$ individual.

We ask the testing participants to rate the recommendation accuracy according to their affective response to the two adjectives "Feminine" and "Technological". Their answers are shown in Figure 10. These results suggest that the recommender has good accuracy in the "Feminine" adjective, averaging 92%. And a lower accuracy in the "Technological" adjective, averaging 64%. This could be due to the participants' high disagreement with this adjective, like what we observe in Figures 7 and 8. The participants with "Neutral openness of personality" tend to have the opposite "technological" affect as people that have "more open" and "less open" personalities.

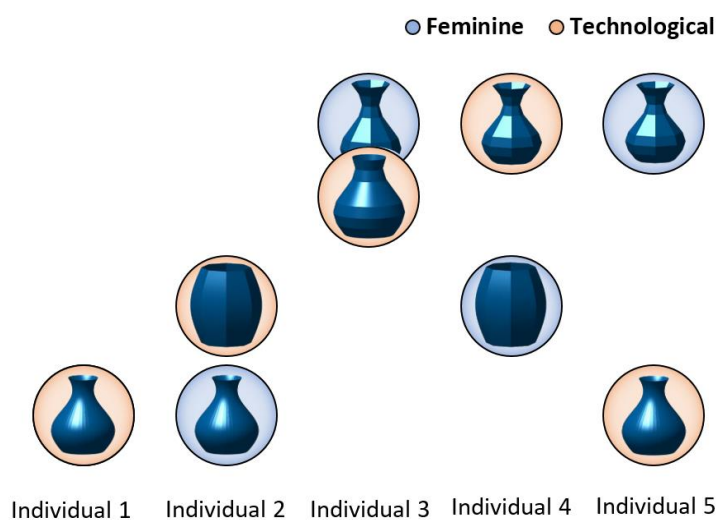

**Figure 9**. Recommended designs for "Feminine" and "Technological" affects

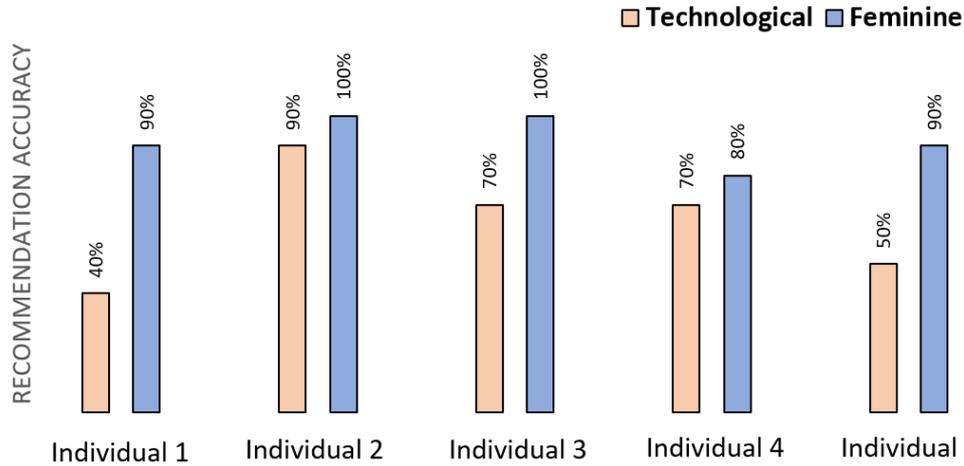

Figure 10. Feedback on the accuracy of the aesthetic design recommendation

## 5 CONCLUSION

This study investigates the problem of aesthetic design recommendations. Using vase design, we suggested an approach based on multiple affective responses and wanted affect. This system recognizes that people have a different affective responses to aesthetic designs and models the individual response to aesthetics in relation to the aspect of openness in personality, plus the element of exposure and the mood during the aesthetic evaluation. The results showed that these individual characteristics are relevant and helped build a representative model that can predict the affective response to a set of designs using the answers to 7 questions.

The recommender part of this system uses the consumer's desired affect and returns the aesthetic design that corresponds to the highest value in that particular affect. Thus, this combination can recommend different designs to different people based on their wanted affect. Also, recommends different designs for the same desired affect, for individuals with different characteristics of openness.

This approach is not limited to vase designs. Because the connection between the individual characteristics of openness and the general aesthetic perception is established in the literature, this can be employed in other product design applications. In the future, further studies will be made at the level of the predictive model in terms of optimizing the weights of the personality questions and adjectives. Also, investigate the connection between the aesthetic features and the affective response in order to build a more flexible design recommender that can generate designs instead of recommending from a limited set of designs.